\documentclass[twocolumn,pra,showpacs,superscriptaddress]{revtex4}
\usepackage{amssymb}
\usepackage{amsmath}
\usepackage{graphicx}
\usepackage{subfigure}
\usepackage{natbib}
\usepackage{epsfig}
\usepackage{amsfonts}
\usepackage{mathrsfs}
\usepackage{ulem}
\usepackage{color}
\usepackage[toc,page,title,titletoc,header]{appendix}
\usepackage{CJK}
\usepackage{graphicx}

\begin{document}

\title{$\mathcal{PT} $-Symmetry in Non-Hermitian Su-Schrieffer-Heeger model with complex boundary potentials}

\author{Baogang Zhu}
\affiliation{Beijing National
Laboratory for Condensed Matter Physics, Institute of Physics,
Chinese Academy of Sciences, Beijing 100190, China}
\author{Rong L\"u}
\affiliation{Department of Physics, Tsinghua University, Beijing 100084, China}
\affiliation{Collaborative Innovation Center of Quantum Matter, Beijing, China}
\author{Shu Chen}
\email{schen@aphy.iphy.ac.cn} \affiliation{Beijing National
Laboratory for Condensed Matter Physics, Institute of Physics,
Chinese Academy of Sciences, Beijing 100190, China}
\affiliation{Collaborative Innovation Center of Quantum Matter, Beijing, China}

\begin{abstract}
We study the parity- and time-reversal ($\mathcal{PT}$) symmetric non-Hermitian Su-Schrieffer-Heeger (SSH) model with two conjugated imaginary potentials $\pm i\gamma $ at two end sites. The SSH model is known as one of the simplest two-band topological models which has topologically trivial and nontrivial phases. We find that the non-Hermitian terms can lead to different effects on the properties of the eigenvalues spectrum in topologically trivial and nontrivial phases. In the topologically trivial phase, the system undergos an abrupt transition from unbroken $\mathcal{PT} $-symmetry region to spontaneously broken $\mathcal{PT} $-symmetry region at a certain $\gamma_{c}$, and a second transition occurs at another transition point $\gamma_{c^{'}}$ when further increasing the strength of the imaginary potential $\gamma$. But in the topologically nontrivial phase, the zero-mode edge states become unstable for arbitrary nonzero $\gamma$ and the $\mathcal{PT}$-symmetry of the system is spontaneously broken, which is characterized by the emergence of a pair of conjugated imaginary modes.

\end{abstract}


\maketitle
\date{today}

\section{Introduction}

One of the fundamental axioms in the Dirac-von Neumann formulation of quantum mechanics is that all physical observables must be represented by Hermitian operators in the Hilbert space \cite{Shankar}, which leads to real energy eigenvalues and guarantees the conservation of probability. However, it is found that a wide class of non-Hermitian Hamiltonians can exhibit an entirely real eigenvalue spectrum if these Hamiltonians have parity-time ($\mathcal{PT} $) symmetry \cite{Bender}. Although whether the non-Hermitian Hamiltonian can define real quantum systems is still debated, intensive efforts have been invested in the study of non-Hermitian Hamiltonians which undergo "spontaneous $\mathcal{PT} $-symmetry breaking" transitions between real and complex eigenvalues (for a recent review, see Ref. \cite{Bender07} and references therein). In order to investigate the physical meaning of a non-Hermitian system with real spectrum, a metric-operator theory method has been presented to map the non-Hermitian Hamiltonian to an equivalent Hermitian Hamiltonian \cite{Mostafazadeh}. Based on these ideas, many $\mathcal{PT} $-symmetric systems have been studied, including quantum field theories \cite{Bender04}, open quantum systems \cite{Rotter09}, the Anderson models for disorder systems \cite{Goldsheid98,Heinrichs01,Molinari09}, the optical systems with complex refractive indices \cite{Klaiman08,Sukhorukov10,Ramezani12,Longhi09,Musslimani08,Luo13}, and the Dirac Hamiltonians of topological insulators \cite{Hu11}. Moreover, some efforts have been made to study the non-Hermitian $\mathcal{PT} $-symmetric discrete system, such as the tight-binding chain \cite{Bendix,Song09,Joglekar,Song10}. In recent years the progress in photonic lattices and photonic crystals have opened up avenues for experimentally verification of these theorems \cite{El-Ganainy,Christian,Guo,Makris,Bittner12,Regensburger12}.

Recent theoretical works have shown the existence of $\mathcal{PT}$-symmetric phases in the one-dimensional (1D) tight-binding chain with conjugated imaginary potentials located at boundary sites if the strength of the boundary potential is smaller than a critical value \cite{Song09}. When the hopping amplitude of the tight-binding chain is modulated alternatively, the model is generally referred as  the Su-Schrieffer-Heeger (SSH) model \cite{Su79}, which was originally proposed to describe the 1D polyacetylene.  Despite of its deceptively simple form, the SSH model shows rich physical phenomena, such as topological soliton excitation, fractional charge and nontrivial edge states \cite{Takayama80,Jackiw76,Su88,Ruostekoski02,Ryu02,Ganeshan13,Chen}, as it serves as a topologically nontrivial prototype model. Due to the rapid advances in topological insulators \cite{TIreview}, the SSH and the extended SSH models have attracted increasing attention as one of the simplest systems of 1D topological insulators \cite{Ryu10,Schnyder08}. It is known that various physical systems can be mapped to the SSH model, such as the two-dimensional graphene ribbon \cite{Delplace11}, the p-orbit ladder-like optical lattice atomic system \cite{Li13}, and the off-diagonal bichromatic 1D system \cite{Ganeshan13}. A main feature of the SSH model is the existence of two topologically different phases which can be distinguished by the presence or absence of two-fold degenerate zero-mode edge states under the open boundary condition (OBC). It is interesting to study how the topologically different phases are affected by the presence of complex boundary potentials under $\mathcal{PT}$-symmetric condition.

To this end, in this work we consider the non-Hermitian $\mathcal{PT} $-symmetric SSH model, which is constructed by adding two conjugated imaginary potentials $\pm i\gamma$ at the end sites of a SSH model under the OBC. In general case, the complex potentials are usually utilized as the non-Hermitian terms to describe physical gain and lose mechanisms phenomenologically \cite{Christian,Guo,Makris}. The whole non-Hermitian Hamiltonian of SSH model is found to possess $\mathcal{PT} $ symmetry despite breaking of $\mathcal{P}$ and $\mathcal{T}$ symmetry separately. We shall focus on the physical effect of conjugated imaginary boundary potentials on the eigenvalues and eigenfunctions of the system in different phases of SSH model.
Our results indicate that the non-Hermitian terms can lead to different behaviors in the topologically nontrivial and trivial phases of SSH model. In the topologically nontrivial phase, the energy spectrum shows complex eigenvalues as long as $\gamma $ is nonzero. While in the topologically trivial phase, the system shows entirely real spectra when $\gamma < \gamma_c $, and when increasing $\gamma$, the system undergos an abrupt phase transition from unbroken $\mathcal{PT} $-symmetry region to spontaneously broken $\mathcal{PT} $-symmetry region at the transition point $\gamma_{c}$. When $\gamma > \gamma_{c}$, the energy spectrum shows 4 complex eigenvalues and the other $2N-4$ eigenvalues remain real, where $2N$ is the total number of sites. As $\gamma $ continues to increase, there exists another transition point $\gamma_{c'}$, above which the bifurcation of imaginary parts of eigenvalues emerges and the complex eigenvalues become purely imaginary.

The paper is organized as follows. In Sec. \uppercase\expandafter{\romannumeral2}, we present the Hamiltonian of the non-Hermitian SSH model with  $\mathcal{PT} $ symmetry. In Sec. \uppercase\expandafter{\romannumeral3}, we study the spectrum of eigenvalues of the non-Hermitian SSH model, and discuss effects of the non-Hermitian terms on properties of the system in topologically trivial and nontrivial phases. Finally, we give the conclusion in Sec. \uppercase\expandafter{\romannumeral4}.

\section{model Hamiltonian}

We consider the 1D non-Hermitian SSH model which describes a tight-binding chain with alternatingly modulated nearest-neighbor hopping parameters and two additional conjugated imaginary on-site potentials at two end sites.
The Hamiltonian can be written as
\begin{eqnarray}
H &=&H_{SSH}+U, \label{Total Hamiltonian}
\end{eqnarray}
where $H_{SSH} $ is the conventional SSH model,
\begin{eqnarray}
&&H_{SSH}  \notag \\
&=&\underset{i=1}{\overset{N}{\sum }}[t(1-\Delta \cos \theta
)c_{2i-1}^{\dagger }c_{2i}+t(1+\Delta \cos \theta )c_{2i}^{\dagger }c_{2i+1}
\notag \\
&&+h.c.],
\end{eqnarray}
with $2N$ the total number of lattice sites. The $ U$ term describes two additional conjugated imaginary on-site potentials acting at the two end sites,
\begin{equation}
U=-i\gamma c_{1}^{\dagger }c_{1}+i\gamma c_{2N}^{\dagger }c_{2N},
\end{equation}
in which particles loss at the 1st site and gain at the $2N$-th site, where $\gamma>0$ is the strength of imaginary potential. $c_{n}^{\dagger }$ ($c_{n} $) is the creation (annihilation) operator on the $n$-th site.  A sketch of the lattice is shown in Fig. 1(a) with hopping parameters given alternatively by $t_{-}$ and $t_{+}$, where $t_{\pm}=t(1\pm \Delta\cos{\theta})$ with $t_{-}$ denoted by the red dashed line, and $t_{+}$ denoted by the green solid line. The parameter $\Delta $ is the dimerization strength and $\theta $ is an introduced tuning parameter, which can vary from $-\pi$ to $\pi$ continuously. 
For convenience, $\Delta$ is defined as $\left\vert \Delta \right\vert <1 $ and $t = 1$ is set as the unit of energy.
\begin{figure}[!htb]
\includegraphics[width=9cm]{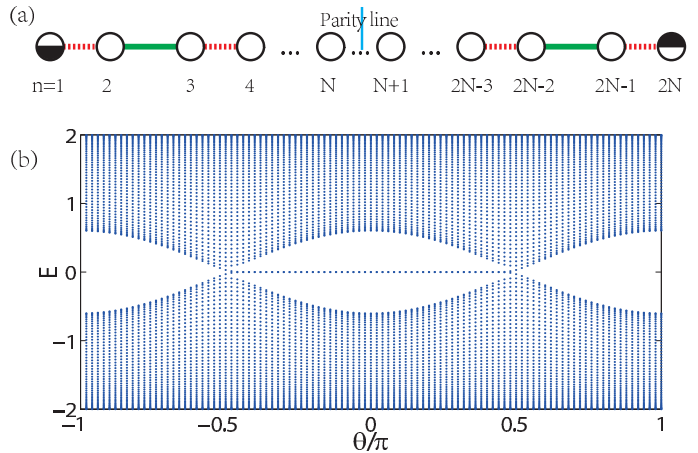}
\caption{(a) Schematic diagram of the SSH model with two additional conjugated imaginary potentials on the 1st and the $2N$-th site. The red dashed line hopping takes value $t(1- \Delta\cos{\theta})$ and the green solid line hopping takes $t(1+ \Delta\cos{\theta})$. (b) Energy spectrum for the conventional SSH model with parameters $\gamma =0$, $\Delta =0.3$, $t=1$, and $2N=100$ under the open boundary condition.}
\end{figure}

In the absence of imaginary boundary potentials, the conventional SSH model is known as the simplest 1D two-band model exhibiting topologically nontrivial properties. As shown in Fig. 1(b) for $\gamma=0$, the SSH model has the topologically nontrivial phase in the regime of $-\pi/2 < \theta < \pi/2$ characterized by the presence of zero-mode edge states under the OBC, whereas no edge states exist in the regimes of $-\pi \leq \theta < -\pi/2$ and $\pi/2 < \theta \leq \pi$ (or equivalently the regime of $-3\pi/2 < \theta < -\pi/2$) corresponding to the topologically trivial phase. For the SSH model under the periodical boundary condition (PBC), these two phases are distinguished by the Berry phase, which takes 0 in the trivial phase and $\pi $ in the nontrivial phase. The zero-mode edge states in the nontrivial phase are topologically protected by both the inversion symmetry and particle-hole symmetry \cite{Ryu02}.

Before we begin with the study of non-Hermitian SSH model, let us list the relevant symmetry properties to be used in this paper. In general, $\mathcal{P}$ and $\mathcal{T}$ are defined as the space-reflection (parity) operator and the time-reversal operator, whose effects are given by $p\rightarrow -p$, $x\rightarrow -x$\, and $p\rightarrow -p$, $x\rightarrow x$, $i\rightarrow -i$, respectively. A Hamiltonian is said to be $\mathcal{PT}$ symmetric if it follows the relation $[\mathcal{PT}, H]=0 $. Furthermore, according to the symmetry of the eigenfunctions \cite{Bender}, the Hamiltonian $H$ can be classified to be either unbroken $\mathcal{PT}$ symmetry or broken $\mathcal{PT}$ symmetry. The time-independent Schr\"odinger equation of eigenfunction $\left\vert \psi \right\rangle $ is given by
\begin{eqnarray}
H\left\vert \psi \right\rangle = E\left\vert \psi \right\rangle, \label{Schrodinger Eq}
\end{eqnarray}
where $E$ is the corresponding eigenvalue. If all the eigenfunctions have $\mathcal{PT}$ symmetry,
\begin{eqnarray}
\mathcal{PT}\left\vert \psi \right\rangle = \left\vert \psi \right\rangle, \label{PTsymmetry}
\end{eqnarray}
then the system has the unbroken $\mathcal{PT}$ symmetry and all the corresponding eigenvalues are real.
But if not all the eigenfunctions obey Eq. (\ref{PTsymmetry}), the system has the broken $\mathcal{PT}$ symmetry, and the eigenvalues of broken $\mathcal{PT}$ symmetry eigenfunctions are complex.

In the discrete lattice case, the effects of $\mathcal{P}$ and $\mathcal{T}$ are $\mathcal{P}c_{i}\mathcal{P}=c_{2N+1-i}$, and $\mathcal{T}i\mathcal{T}=-i$, respectively. For the non-Hermitian SSH model studied in this paper, we can show that
\begin{eqnarray}
\mathcal{P}H_{SSH}\mathcal{P}=H_{SSH} &,& \mathcal{T} H_{SSH}\mathcal{T}=H_{SSH}\nonumber\\
\mathcal{P}U\mathcal{P}=-U &,& \mathcal{T}U\mathcal{T}=-U\nonumber\\
\mathcal{P}H\mathcal{P}\neq H,\mathcal{T}H\mathcal{T} &\neq& H,\mathcal{PT}H\mathcal{TP}=H.
\end{eqnarray}
Then the Hamiltonian $H$ in Eq. (\ref{Total Hamiltonian}) for the non-Hermitian SSH model has neither $\mathcal{P}$ nor $\mathcal{T}$ symmetry separately, but $H$ is invariant under their combined operation $\mathcal{PT}$.

\section{Results and Discussions}

In this section, we present numerical calculations of Schr\"odinger equation Eq.(\ref{Schrodinger Eq}) of the non-Hermitian $\mathcal{PT}$ symmetric SSH model under the OBC, and investigate effects of two conjugated imaginary boundary potentials on the energy spectrum of the system. As shown in Fig. 1(b), eigenvalue spectrum of the conventional SSH model in regimes of $-\pi/2 < \theta < \pi/2$ and $-3\pi/2 < \theta < -\pi/2$ show different features, corresponding to topologically nontrivial and trivial phases. Taking the boundary terms $\pm i\gamma$ into account, we may expect that imaginary boundary potentials have different effects on the eigenvalue spectrum in these two different regimes.

We first consider the topologically nontrivial regime, i.e., the regime of $-\pi/2 <\theta<\pi/2 $. In Fig.2 we show the real and imaginary parts of the eigenvalues of the system under different conditions. For the case of weak imaginary boundary potentials with $\gamma=0.1$, one can observe that the real part of eigenvalues shown in Fig. 2(a) has a similar structure to that of the conventional SSH model shown in Fig. 1(b), i.e., there still exist mid-gap modes with $Re(E)=0$ in this regime, which may be viewed as a reminiscent of the zero mode. Checking eigenvalues in this regime, we find that there are only two complex eigenvalues with the form of $\pm ib$ (here $b$ is a function of $\theta$ and $\gamma$), i.e., the spectrum of the system is composed of two conjugated imaginary eigenvalues and $2N-2$ real eigenvalues. For the case with a smaller $\gamma$, eigenvalues of the system have a similar structure to that shown in Fig. 2(a). Actually we find that the imaginary part of energy spectrum emerges in the whole regime of  $-\pi/2 <\theta<\pi/2 $ once $\gamma$ is nonzero. This observation indicates that the zero-mode edge states of the SSH model in the regime of $-\pi/2 <\theta<\pi/2 $ become unstable for an arbitrary nonzero $\gamma$ and meanwhile the $\mathcal{PT}$ symmetry of the system is spontaneously broken \cite{Hu11,Esaki11}. As $\gamma$ continues to increase to other values, e.g., $\gamma=0.8$, $1$, $2$ and $3$, the spectrum in the regime of $-\pi/2 <\theta<\pi/2 $ has the similar structure, i.e., there exist only a pair of conjugated imaginary eigenvalues, as shown in Fig 2.(b)-(e).

\begin{figure}
\includegraphics[width=9cm]{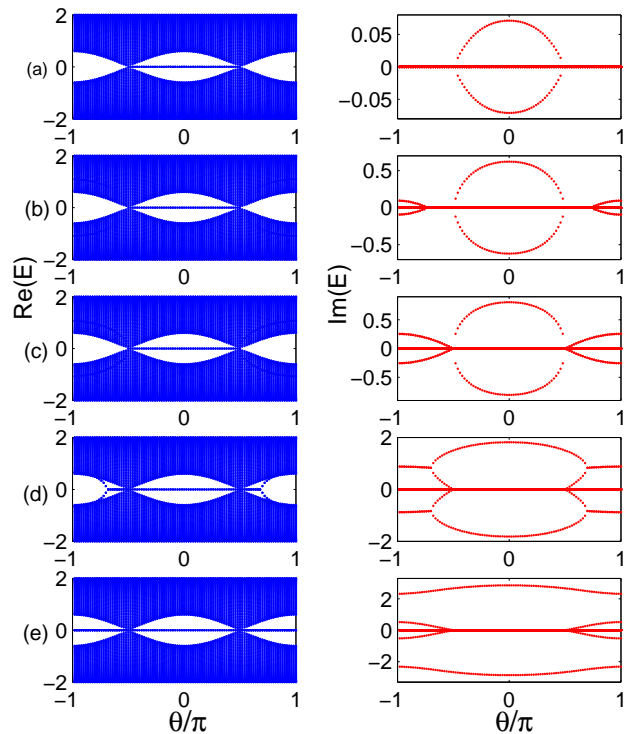}
\caption{The real and the imaginary part of the eigenvalue spectrum of the non-Hermitian SSH model as a function of $\theta$ with parameters $\Delta =0.3$, $t=1$, and $2N=100$ for different $\gamma$: (a) $\gamma=0.1$, (b) $\gamma=0.8$, (c) $\gamma=1$, (d) $\gamma=2$ and (e) $\gamma=3$ under the open boundary condition. Left figures represent the real part of the spectrum and right figures represent the imaginary part.}
\end{figure}

Another interesting observation in this regime is that the imaginary eigenvalue $Im(E)$ takes its maximum at $\theta=0$ and its minimum at the boundary of $\theta=\pm\pi/2$ for a given $\gamma$. For $-\pi/2 < \theta <\pi/2$, we have $t_{-}<t_{+} $, i.e., the hopping amplitude denoted by the red dashed line in Fig. 1(a) is weaker than that denoted by the green solid line, which leads to the formation of edge states for the conventional SSH model under the OBC. The more close of $\theta$ to 0, the bigger of the ratio $t_{+}/t_{-}$ is, and then the edge states localized at two end sites are more easily influenced by the imaginary boundary terms. In the extreme case with $\Delta=1$ and $\theta=0$, we have $t_{+}=2$ and $t_{-}=0 $, which implies the 1st site and the $2N$-th site being completely isolated from the other sites of the 1D chain. So a pair of imaginary modes with the imaginary eigenvalues of $\pm i \gamma$ emerge once nonzero $\gamma$ is fixed on.

\begin{figure}
\includegraphics[width=9cm]{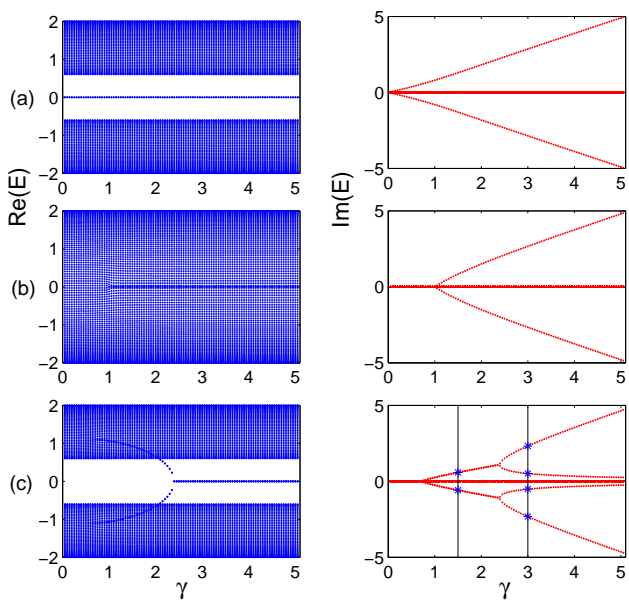}
\caption{The real (left) and imaginary (right) parts of the eigenvalue spectrum versus $\gamma$ for the system in different topological regimes. (a) system in the topologically nontrivial regime with $\theta=0$; (b) system on the phase boundary with $\theta=-\pi/2$; (c) system in the topologically trivial regime with $\theta=-\pi$. Other parameters are the same as in Fig. 2. The conjugated imaginary eigenvalues marked by the star points correspond to systems with $\gamma=1.5$ and $3$, respectively.}
\end{figure}

\begin{figure*}
\includegraphics[width=6in]{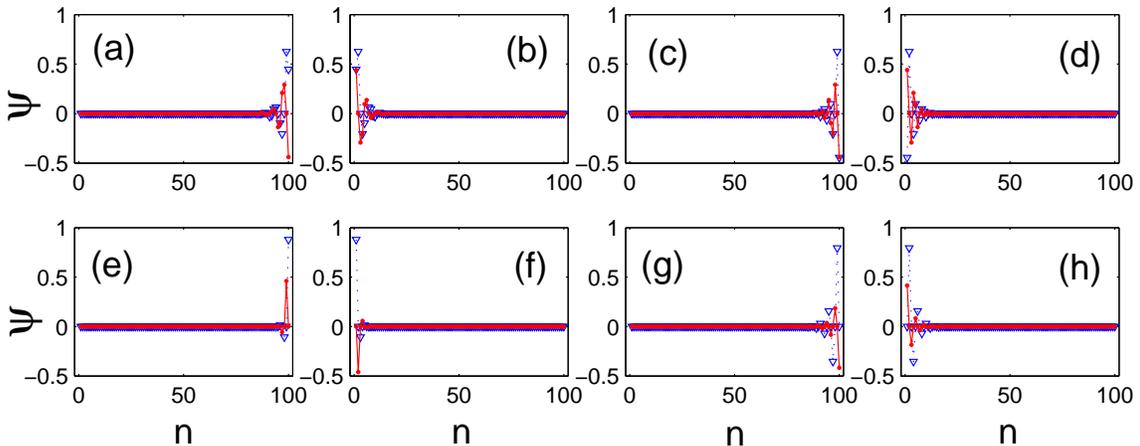}
\caption{Distributions of eigenfunctions for systems with the complex eigenvalues marked in Fig.3(c). (a)-(d) are for the system with $\gamma=1.5$ corresponding to eigenvalues $0.925+0.587i$ (in (a)), $0.925-0.587i$ (in (b)), $-0.925+0.587i$ (in (c)), and $-0.925-0.587i$ (in (d)), respectively. Distributions of eigenfunctions for for system with $\gamma=3$ are shown in (e)-(h) corresponding to eigenvalues $2.319i$ (in (e)), $-2.319i$ (in (f)), $0.517i$ (in (g)) and $-0.517i$ (in (h)), respectively. The blue dashed line stands for the real part of the eigenfunction and the red solid line stands for the imaginary part.}
\end{figure*}

Next we consider the topologically trivial regime, i.e., the regime of $-3\pi/2 < \theta < -\pi/2$. As shown in Fig.2, one can observe that eigenvalues in this regime display obviously different behaviors from those in the regime of  $-\pi/2 < \theta < \pi/2$. It turns out that the system shows much richer features in this regime when increasing the strength of imaginary potential.
In the case of weak imaginary potential, e.g., $\gamma=0.1$ as shown in Fig. 2(a), the non-Hermitian SSH model has an entirely real eigenvalue spectrum, which indicates that the $\mathcal{PT}$ symmetry is unbroken in the presence of weak conjugated imaginary boundary potentials.
When increasing the strength of boundary potentials $\gamma$, one may expect that complex eigenvalues would occur in this regime if $\gamma$ is larger than a critical value of $\gamma_{c}$. Here $\gamma_{c}$ is associated with $\theta $ and thus can be defined as $\gamma_{c,\theta}$. We find that complex eigenvalues begin to emerge when $\gamma>0.7$ for the system with $\theta=\pm \pi$, i.e., $\gamma_{c,\pm \pi} = 0.7$. With $\gamma $ continuously increasing, larger regimes near $\theta=\pm \pi$ display complex eigenvalues. As an example, the energy spectrum of the system with $\gamma=0.8$ is shown in Fig. 2(b). One can clearly see that regimes near $\theta=\pm \pi$ have complex eigenvalues corresponding to the breaking of $\mathcal{PT}$ symmetry. On the other hand, eigenvalues near $\theta=\pm \pi/2$ are entirely real,
and thus the system still has unbroken $\mathcal{PT}$ symmetry.

As $\gamma$ continues to increase to 1, the system with $\theta=\pm \pi/2$ displays complex eigenvalues, i.e., the critical value
of $\gamma$ is $\gamma_{c,\pm \pi/2}=1$. As shown in Fig. 2(c), in this case complex eigenvalues emerge in the whole regime of $\theta$, which indicates that the $\mathcal{PT}$ symmetry of systems in the whole regime is spontaneously broken.
Particularly, in the case of $\theta=\pm\pi/2 $, the Hamiltonian of the non-Hermitian SSH model reduces to
\begin{eqnarray}
H=t\underset{i=1}{\overset{2N}{\sum }}(c_{i}^{\dagger}c_{i+1}+H.c.)-i\gamma c_{1}^{\dagger }c_{1}+i\gamma c_{2N}^{\dagger }c_{2N},
\end{eqnarray}
and the system turns to an isotropy tight-binding chain with $t_{+}=t_{-}$, which can be solved analytically \cite{Song09} with the critical point given by $\gamma_{c}=1$, which agrees with our results in this specific case. Fig. 2(c) also indicates that in the regime of $-3\pi/2 <\theta< -\pi/2$, the imaginary eigenvalue $Im(E)$ takes its maximum at $\theta=-\pi$ and approaches zero as $\theta \rightarrow -\pi/2$ for $\gamma=1$.

When $\gamma >1$, a new phase emerges in the topologically trivial regime as $\theta$ close to $\theta=\pm \pi/2$, which can be characterized by the existence of the bifurcation of the imaginary parts of the eigenvalues, i.e., one pair of conjugated imaginary values splits into two pairs. An example of $\gamma=2$ is given in Fig. 2(d), which shows that the spectrum is composed of $4$ imaginary and $2N-4$ real eigenvalues in the regime of $-0.695 \pi  < \theta<-\pi/2$. When $\gamma>2.39$, the phase with four imaginary eigenvalues appears in the whole regime of $ - 3 \pi/2  < \theta<-\pi/2$, which is shown in Fig. 2(e) for the example system with $\gamma=3$.

The above results indicate the existence of three different phases in the regime of $-3\pi/2 < \theta< -\pi/2$, whereas there exists only one phase in the regime of $-\pi/2 < \theta <\pi/2$. To see clearly how the spectrum changes in different phases with the increase in $\gamma$, we show the real and imaginary parts of eigenvalues as a function of $\gamma$ for the system in different topological regimes. Typical examples are displayed in Fig. 3(a), (b) and (c) by assigning $\theta$ as $0$, $-\pi/2$ and $-\pi$, respectively. It is clear that in the topologically nontrivial regime the imaginary part of eigenvalues emerges once $\gamma$ is nonzero as shown in Fig. 3(a), whereas the imaginary part of eigenvalues begins to appear at $\gamma_{c,-\pi/2}=1$ at the boundary line of $\theta=-\pi/2$. In the topologically trivial regime with $\theta=-\pi$, the system has a purely real spectrum when $\gamma<0.7$ and undergos a phase transition at $\gamma_{c,-\pi}=0.7$. When $\gamma>\gamma_{c,-\pi}$, 4 complex eigenvalues emerge with the form of $\pm a \pm i b$ ($a,b$ are functions of $\theta$ and $\gamma$), and the other $2N-4$ eigenvalues are real. With further increasing $\gamma$, the structure of the spectrum changes at a second transition point $\gamma_{c^{'},-\pi}=2.39$, at which these four complex eigenvalues become purely imaginary, as shown in Fig. 3(c). Particularly, in the limit of $\gamma \rightarrow \infty$, the values of a pair of the conjugated imaginary modes tend to zero, which suggests the emergence of a pair of zero modes in this limit. When $\theta$ deviates from $\pm \pi$ but still in the topologically trivial regime, the spectrum versus $\gamma$ has similar structures to that of the case of $\theta=- \pi$. As the parameter $\theta$ approaches to the phase boundary point of $\theta=\pm \pi/2$, the transition points $\gamma_{c,\theta}$ and $\gamma_{c^{'},\theta}$ approach to $1$. At the boundary of $\theta=\pm \pi/2$, the two transition points merge to one, i.e., $\gamma_{c,\pm \pi/2}=\gamma_{c^{'},\pm \pi/2}=1$, as shown in Fig. 3(b).

The change of the structure of spectrum as a function of $\gamma$ has revealed different phases of the system with $\theta=-\pi$. While all eigenstates in the regime of $\gamma<\gamma_{c}$ are real and spread over the whole lattice, complex modes emerge in the regime of $\gamma>\gamma_{c}$ corresponding to the existence of complex eigenvalues. To see how the complex eigenstates distribute on the lattice, we display distributions of eigenfunctions in Fig. 4 corresponding to four complex eigenvalues for systems with $\gamma=1.5$ and $\gamma=3$, respectively, which are marked in Fig. 3 (c). As shown in the Fig. 4, eigenfunctions corresponding to the conjugated complex eigenvalues, e.g., $0.925 \pm 0.587i$, $-0.925 \pm 0.587i$, $\pm 2.319i$ and $\pm 0.517i$, are located at the right and left boundaries of the lattice, respectively, whereas eigenfunctions of the other $2N-4$ real eigenvalues are all bulk states. Before ending the paper, we would like to remark that, while our results are obtained by solving the eigen-equation directly, it would be also interesting to study the scattering problem for transmission through the system on different regimes by using the S-matrix method \cite{Rotter09,Rotter03} in the future work.


\section{Summary}

In summary, we have studied the $\mathcal{PT}$-Symmetry of the non-Hermitian SSH model with two additional conjugated imaginary on-site potentials at the two end sites. Our results indicate that the non-Hermitian boundary potential terms lead to different behaviors in two different phase regimes of the SSH model. In the topologically nontrivial phase regime, the model has complex eigenvalues once the strength of imaginary potential is nonzero. However, in the topologically trivial phase regime, this model exhibits an unbroken $\mathcal{PT} $-symmetry phase with a real eigenvalue spectrum for $\gamma < \gamma_{c}$, and a spontaneous-$\mathcal{PT} $-symmetry-broken phase with $2N-4$ real and 4 complex eigenvalues for $\gamma > \gamma_{c}$. When increasing the strength of imaginary potential further, the model has another transition point $\gamma_{c^{'}}$, above which the bifurcation of imaginary parts of eigenvalues appears and four complex eigenvalues become purely imaginary.

\section*{Acknowledgments}

This work has been supported by National Program for Basic Research of MOST, by NSF of China under Grants No.11374354, No.11174360, and No.11121063, and by the Strategic Priority Research Program of the Chinese Academy of Sciences under Grant No. XDB07000000. R. L. is supported by the NSFC under Grant No. 10974112 and the National Basic Research Program of China (973 Program) Grant No. 2011CB606405 and No. 2013CB922000.

\end{document}